\documentclass[a4paper,11pt]{article}

\usepackage{pos}
\usepackage{amsmath,lineno}

\newcommand{\dif}{\mathrm{d}}

\title{Observational constraints on cosmic-ray escape from ultra-high energy accelerators}
\ShortTitle{Observational constraints on cosmic-ray escape from UHE accelerators}

\author*[a]{Antonio Condorelli}
\author[b]{Quentin Luce}
\author[a]{Sullivan Marafico}
\author[a]{Jonathan Biteau}
\author[a]{Olivier Deligny}

\affiliation[a]{Université Paris-Saclay, CNRS/IN2P3, IJCLab,\\
   Orsay, France,}

\affiliation[b]{Karlsruhe Institute of Technology, Institute for Experimental Particle Physics (ETP),\\
Karlsruhe, Germany}

\emailAdd{condorelli@ijclab.in2p3.fr}

\abstract{The energy spectrum and mass composition of ultra-high energy cosmic rays inferred at the Pierre Auger Observatory are used to derive a benchmark scenario for the emission mechanisms at play in extragalactic accelerators as well as for their energetics and for the abundances of elements in their environments. Assuming a distribution of sources following the density of stellar mass, the gradual increase of the cosmic ray mass number observed on Earth from $\simeq$2\:EeV up to the highest energies is shown to call for nuclei accelerated up to an energy proportional to their electric charge and emitted with a hard spectral index. In addition, the inferred flux of protons down to $\simeq$0.6\:EeV is shown to require for this population a spectral index significantly softer than that of heavier nuclei. This is consistent with in-source interactions that shape the energy production rate of injected charged nuclei differently from that of the secondary neutrons escaping from the confinement zone. Together with the inferred abundances of nuclei, these results provide constraints on the radiation levels in the source environments. Within this scenario, an additional component that falls off steeply with increasing energy up to the ankle feature is necessary to make up the all-particle flux in the sub-ankle energy range.}

\FullConference{%
  *** 27th European Cosmic Ray Symposium - ECRS ***\\
  *** 25-29 July 2022 ***\\
  *** Nijmegen, the Netherlands ***
}

\begin{document}
\maketitle

\section{Introduction} 
\label{sec:intro}

The origin of ultra-high energy cosmic rays (UHECRs) is currently unknown. Their identification is based mostly on catching a potentially interesting class of astronomical objects in the UHECR arrival directions. An observation remains remains elusive, although new investigations have allowed for broad statements to be made. Above $\simeq 8$ EeV \citep{PierreAuger:2017pzq}, an anisotropy at large scales has been discovered, the intensity and direction of which are consistent with assumptions drawn from sources distrubuted in a similar way to extragalactic matter \citep{PierreAuger:2017pzq,PierreAuger:2018zqu}. A correlation between UHECR arrival directions and the flux distributions of massive, star-forming, or active galaxies within 200 Mpc provides evidence for anisotropy at higher energies ($\gtrsim 40$ EeV) \citep{PierreAuger:2018qvk,PierreAuger:2022rfz}. Overall, these results suggest that UHECRs are primarily of extragalactic origin, at least above the so-called ankle energy ($\simeq 5$ EeV).

The observed energy spectrum and chemical composition of UHECRs on Earth are the result of emission processes that include acceleration mechanisms, losses and escape from source environments, as well as propagation effects. Differently from, and complementing to anisotropies, these two observables provide constraints, helping to infer the features of the acceleration processes, the energetics of the sources, and the abundances of elements in the source environments. Following this idea, various investigations have set the groundwork for a generic scenario that broadly replicates the observations. \citep{Aloisio:2013hya,PierreAuger:2016use,PierreAuger:2021mmt}.

The intensity of the different nuclear components at the sources, generally assumed as standard candles and uniformly distributed in a comoving volume, is supposed to drop off at the same magnetic rigidity to explain the steady increase with energy of mass number $A$ observed on Earth \citep{PierreAuger:2010ymv,PierreAuger:2014gko}. This agrees to the underlying assumption that electromagnetic processes accelerate particles to a maximum energy proportional to their electric charge $Z$. The abundance of nuclear elements is observed to be dominated by intermediate-mass ones, ranging from He to Si, accelerated to $E^{Z}_{\mathrm{max}}\simeq 5Z\,\text{EeV}$ and leaving from the source environments with a very hard spectral index $\gamma$.

The constraints on the characteristics of the sources are actually discussed in \citep{Luce_2022}. To accomplish so, data from the Pierre Auger Observatory beyond 0.63 EeV ($10^{17.8}$ eV) were taken into account. However, in contrast to other approaches, only the proton spectrum is used in the energy range between 0.63 EeV and 5 EeV. This method allows us to reconstruct the extra-galactic component without introducing any ad hoc nuclear components at the lowest energies, to represent what is generally believed to be the upper end of the galactic component, in the combined fit of both energy spectrum and mass composition. Because of both observational systematics and theoretical uncertainties, such modeling alters the reconstruction of the extragalactic proton component with biases intrinsic in the choices made.

\section{Generic model of UHECR production with in-source interactions} 
\label{sec:model}

The main assumptions of the benchmark astrophysical model employed here can be found in  \cite{PierreAuger:2016use}. The non-thermal processes responsible for accelerating the different types of particles are modeled using power-law spectra proportional to the charge $Z$, such as $E^{Z}_{\mathrm{max}}=ZE_{\mathrm{max}}$, while an exponential suppression is used to describe the end of the acceleration process in the absence of any strong indication from theory. The sources are believed to accelerate diverse proportions of nuclei, which are represented by five stable ones: hydrogen ($^1$H), helium ($^4$He), nitrogen ($^{14}$N), silicon ($^{28}$Si), and iron ($^{56}$Fe).
The ejection rate per comoving volume unit and per energy unit of nucleons is usually modeled as
\begin{equation}
\label{eqn:qN}
    q_{\rm p}(E) = q_{0{\rm p}}\left(\frac{E}{E_0}\right)^{-\gamma_{\rm p}}f_{\mathrm{supp}}(E,Z_{\rm p}),
\end{equation}

Using $Z_{\rm p}=1$,which is a single ejection rate $q_{0{\rm p}}$, spectral index $\gamma_{\rm p}$, and suppression function $f_{\mathrm{supp}}(E,Z_{\rm p})$ for both escaping protons and neutron decay protons. In this case, $E_0$ is arbitrarily fixed to 1\:EeV. Generically, the ejection rate of nuclei with mass number ${A}_i$ is likewise modeled as
\begin{equation}
\label{eqn:qA}
    q_{{A}_i}(E) = q_{0{A}_i}\left(\frac{E}{E_0}\right)^{-\gamma_{A}}f_{\mathrm{supp}}(E,Z_{{A}_i}),
\end{equation}
For helium, nitrogen, silicon, and iron, there is a single spectral index $\gamma_{A}$ and four separate reference ejection rates $q_{0{A}_i}$. The suppression function used for nucleons and nuclei is the same as in the reference case of \cite{PierreAuger:2016use}.
\begin{equation}
\label{eqn:fsupp}
    f_{\mathrm{supp}}(E,Z) = 
    \begin{cases}
    1 & \mathrm{if~}E\leq E^{Z}_{\mathrm{max}},\\
    \exp{\left(1-E/E^{Z}_{\mathrm{max}}\right)} & \mathrm{otherwise}.
    \end{cases}
\end{equation}
The maximum acceleration energy is expected to be proportional to the electric charge of each element, $E^{Z}_{\mathrm{max}} = ZE_{\mathrm{max}}$, with the five species having a single free parameter $E_{\mathrm{max}}$.

The ejection rate for protons in this approach accounts for both the accelerated ones up to $E_{\mathrm{max}}$ and those produced by escaping neutrons, resulting in an energy of the order of $E/A$ inherited from the nuclei of energy $E$. The maximal energy reached by this population of secondary protons is then $E^{Z}_{\mathrm{max}}/A\simeq E_{\mathrm{max}}/2$, which is not represented in the rigidity-acceleration scheme modeled by Equation~\eqref{eqn:fsupp}. Replacing $E_{\mathrm{max}}$ with $E_{\mathrm{max}}/2$ in $q_{\mathrm{p}}(E)$ would test the extreme situation in which all ejected protons are photodissociation by-products. Modeling of sources and their environments is required for further characterization of the balance between the population of accelerated protons from the initial abundance in the source environment and that of secondaries from nuclear cascades.

For each species $j$, the differential energy production rate per comoving volume unit of the sources, which is directly connected to their differential luminosity, is consequently $\ell_j(E,z)= E^2q_j(E)S(z)$, where $S(z)$ reflects the redshift evolution of the UHECR luminosity density. For convenience, the quantity $\ell_j(E,z)$ is referred to as the differential energy production rate. The bolometric energy production rate per comoving volume unit at redshift $z$, on the other hand, is calculated as $\mathcal{L}_j(E, z)=S(z)\int_{E}^{\infty}\dif E'E'q_j(E')$. We report its average value in a volume spanning $z_\text{min}-z_\text{max}$ as $\bar{\mathcal{L}}_j(E) = \int_{z_\text{min}}^{z_\text{max}} \dif z \left|\frac{\dif t}{\dif z}\right| \mathcal{L}_j(E, z) / \int_{z_\text{min}}^{z_\text{max}} \dif z \left|\frac{\dif t}{\dif z}\right|$, where $t(z)$ is the lookback time.

The density of baryonic matter is used to trace the evolution of the UHECR luminosity density throughout cosmic time. This is supposed to follow star mass density, which is well approximated by a constant out to redshift $z=1$ \citep[e.g.][]{2014ARA&A..52..415M}. As in the benchmark case of \cite{PierreAuger:2016use}, such a constant evolution is expected to hold to a first approximation of $z_\text{max} = 2.5$. The local Universe, on the other hand, has an overdensity because the Milky Way, like most galaxies, is part of a group of galaxies which is embedded in the Local Sheet \citep{2014MNRAS.440..405M}. We use the overdensity correction factor calculated by ~\cite{2019ApJ...872..148C}, which is defined as a function of distance $r$ and is effective up to 30\:Mpc,
\begin{equation}
\label{eqn:rho}
    \frac{\delta\rho(r)}{\bar \rho} = 1+\left(\frac{r}{r_0}\right)^{-\alpha},
\end{equation}
with $r_0=5.4\:$Mpc and $\alpha=1.66$. The evolution of the UHECR luminosity density is then represented by the equation  $S(z) = \delta\rho/\bar \rho$, with $z_\text{max} = 2.5$ and $z_\text{min} = 2\times 10^{-4}$.
The minimal redshift, which corresponds to the Local Group limit ($r\simeq 1\:$Mpc), precludes any divergence in Equation~\eqref{eqn:rho} and effectively removes very-closeby galaxies that would otherwise dominate the UHECR sky, which contradicts observations \cite{PierreAuger:2022rfz}.

In this scenario, we are neglecting the effect of extra-galactic magnetic field on particle propagation, therefore this can thus be termed one-dimensional.

The current all-particle energy spectrum $J(E)$ derives from the integration of all sources' contributions across lookback time, with redshift playing a role:
\begin{equation}
\label{eqn:model}
    J(E) = \frac{c}{4\pi}\sum_{A,A'} \iint \dif z\,\dif E' \left|\frac{\dif t}{\dif z}\right|S(z) q_{A'}(E')\frac{\dif\eta_{AA'}(E,E',z)}{\dif E}.
\end{equation}
The relationship between cosmic time and redshift is derived from the cosmological concordance model, $(\dif t/\dif z)^{-1}= -H_0(1+z)\sqrt{\Omega_{\mathrm{m}}(1+z)^3+\Omega_{\Lambda}}$, where $H_0=70\:$km\:s$^{-1}$\:Mpc$^{-1}$ is the Hubble constant at present time, $\Omega_{\mathrm{m}}\simeq0.3$ is the density of matter (baryonic and dark matter)
and $\Omega_{\Lambda}\simeq0.7$ is the dark-energy density; $\eta_{AA'}(E,E',z)$ is the amount of particles detected on Earth with energy $E$ and mass number $A$ released by sources with energies $E'>E$ and mass numbers $A'>A$ and describes the energy losses and spallation processes. 
In practice, for a given redshift $z_0$ source emitting a nuclear species $A_0$ at energy $E_0$, the corresponding $\eta_{AA'}(E,E',z)$ function is tabulated in bins of $A'$ and $E'$ by propagating a large number of emitted particles (O($10^7$) particles) by using the SimProp package~\citep{Aloisio:2012wj}. 
By repeating the simulations for different $z_0$, $A_0$, and $E_0$ values, the entire $\eta_{AA'}(E,E',z)$  function is tabulated as a 5D histogram, returning the fractions sought. SimProp accounts for pair production, photo-pion production, and photodissociation off the photon fields of interest, which are those from cosmic-microwave-background (CMB) radiation and infrared photons from extragalactic background light (EBL), which is the radiation produced in the Universe since the formation of the first stars. The CMB radiation can be described as  a black-body spectrum with a redshift-dependent temperature $T(z)=T_0(1+z)$, with $T_0=2.725\:$K. The EBL is less well known, particularly in the far infrared and at high redshifts. For the benchmark scenario explored below, we utilise the model of ~\cite{Gilmore:2011ks}, which is consistent with the minimal intensity level derived from galaxy counts as well as indirect measurements derived from gamma-ray absorption at multi-TeV energies \citep[e.g.][for a recent review]{2021arXiv211205952P}.
Equation~\eqref{eqn:model} can be used, with these ingredients, to evaluate the all-particle flux from the various contributions of each individual nuclear component on the condition of assigning values to the five ejection rates $q_{0A_i}$, the two indices $\gamma_{\rm p}$ and $\gamma_A$, and the maximum energy $E_{\rm max}$. These eight parameters are fitted to the data applying the method described in the following section.
\section{Combined fit to energy-spectrum and mass-composition data} 
\label{sec:data}

The expected spectrum represented by Equation~\eqref{eqn:model}  is dependent on various unknown features that characterize acceleration processes, source surroundings, and source energetics. Constraints from the measured energy spectrum and mass composition can contribute in inferring these properties.
We will use, on the one hand, the all-particle energy spectrum inferred from these data ~\citep{PierreAuger:2021hun}, and on the other, the distributions of the slant depth of maximum of shower development (hereinafter called $X_{\mathrm{max}}$), which is the best up-to-date proxy of the primary mass of the particles ~\citep{PierreAuger:2014gko,Bellido:2017cgf} . Using the hadronic-interaction generators to represent the development of the showers, the $X_{\mathrm{max}}$ distributions allow for statistical inference of the energy-dependent mass composition. Here, two hadronic-interaction generators are considered: EPOS-LHC ~\citep{Pierog:2013ria} for the benchmark scenario and Sibyll2.3c~\citep{riehn2017hadronic} as an alternative, both of which are the most recent and best describe the data. As a result, by combining the all-particle energy spectrum and the abundances of the various mass components $J_j(E)$, the different elemental spectra as a function of energy can be calculated.
The energy threshold here taken into account is the nominal one of the $X_{\mathrm{max}}$ detection mode at ultra-high energies, namely ${\simeq}\: 0.63\:$EeV ($10^{17.8}\:$eV). This threshold, incidentally, allows us to investigate the energy spectrum of the ankle feature.
A Gaussian likelihood fit is used to establish the best match between the observed spectrum and that expected from Equation~\eqref{eqn:model} from a set of proposed parameters $\mathbf{\Theta}$. Assuming that the transition to extragalactic UHECRs has already occurred above 5\:EeV,\footnote{ We evaluated that a moderate increase in the value at which the transition to extragalactic UHECRs is considered to be complete would not have any effect on the final results.} the equivalent probability term, $L_J$, is calculated above that threshold. The model is fitted to the $X_\mathrm{max}$ data using the multinomial distribution, which represents how likely it is to observe $k_{mx}$ events out of $n_m$ with probability $p_{mx}$ in each $X_{\mathrm{max}}$ bin in each energy bin $m$. The probabilities $p_{mx}$ are calculated by modeling the $X_{\mathrm{max}}$ distributions in terms of the parameters $\mathbf{\Theta}$ using generators of hadronic interactions. The corresponding likelihood, $L_{X_{\mathrm{max}}}$, is considered as the product of energy bins greater than 5\:EeV. The last contribution to the likelihood comes from the sub-ankle proton component, $L_{J_{\mathrm{p}}}$, which is calculated by weighting the all-particle spectrum below 5\:EeV with the proton abundance $f_{\mathrm{p}}(E)$ from the $X_{\mathrm{max}}$ distributions. The statistical uncertainties in $\hat{J}_{\mathrm{p}}(E)$ are dominated by those in $f_{\mathrm{p}}(E)$, and they are accounted for using a Gaussian likelihood fit.

As a result, the model likelihood is provided by $L=L_JL_{X_{\mathrm{max}}}L_{J_{\mathrm{p}}}$. The goodness-of-fit is measured using a deviance, $D$, which is defined as the negative log-likelihood ratio between a given model and the saturated model that represents the data:
\begin{equation}
\label{eqn:deviance}
D=-2\log{\frac{L_J}{L_J^{\mathrm{sat}}}}-2\log{\frac{L_{X_{\mathrm{max}}}}{L_{X_{\mathrm{max}}}^{\mathrm{sat}}}}-2\log{\frac{L_{J_{\mathrm{p}}}}{L_{J_{\mathrm{p}}}^{\mathrm{sat}}}}.
\end{equation}
The three different contributions are referred to as $D_J$, $D_{X_{\mathrm{max}}}$ and $D_{J_{\mathrm{p}}}$ respectively.

\section{Results} \label{sec:results}
Fitting the model to the data using the benchmark scenario provided in Section~\ref{sec:model}, with free spectral indices for both nucleons and nuclei ($\gamma_{\rm p}\neq\gamma_{A}$), provides the parameters and deviance shown in Table 1 of \cite{Luce_2022} using EPOS-LHC to interpret $X_{\mathrm{max}}$ data. Table 1 of \cite{Luce_2022} includes the results predicted under the assumptions of a proton component dominated by neutron escape (proton maximum energy of $E_{\rm max}/2$), no local overdensity (widely used uniform distribution), and a shared spectral index across the five species ($\gamma_{\rm p}=\gamma_{A}$).
The value of $E_{\mathrm{max}}$, derived by the drop in the nuclear components at $E^Z_{\mathrm{max}}$, indicates that the suppression of the spectrum is due to a combination of the cut-off energy at the sources for the heavier nuclei and energy losses \emph{en route}, as found in \cite{PierreAuger:2017pzq}. The nuclei's spectral index, $\gamma_A$, is defined by the increase in average mass with energy, which is virtually monoelemental, in order to mimic the $X_\mathrm{max}$ distributions as as accurately as possible. The fit's solution thus consists on setting a hard index for nuclei so that the contributions of each element mix as little as possible: high-energy suppression imposed by the cut-off beyond $E^{Z}_{\mathrm{max}}$  and low-energy suppression via the hard index $\gamma_{A}$. This effect, however, does not apply to protons, which persist in an energy range where a mixing of elements is required. The best-fit value of $\gamma_{\mathrm{p}}$ is significantly softer than that of $\gamma$. The addition of $\gamma_{\mathrm{A}}$ improves the fit of the data down to 0.63\:EeV, with a total deviance $D=236.8$ compared to $862.7$ in the case of $\gamma_{\mathrm{p}}=\gamma_{\mathrm{A}}$, therefore the addition of this extra free parameter is sufficiently justified. On the other hand, using local overdensity to trace the source distribution improves the deviation significantly while it has a minor impact on the best-fit parameters. 

\begin{figure}[t]
\centering
\includegraphics[scale=.5]{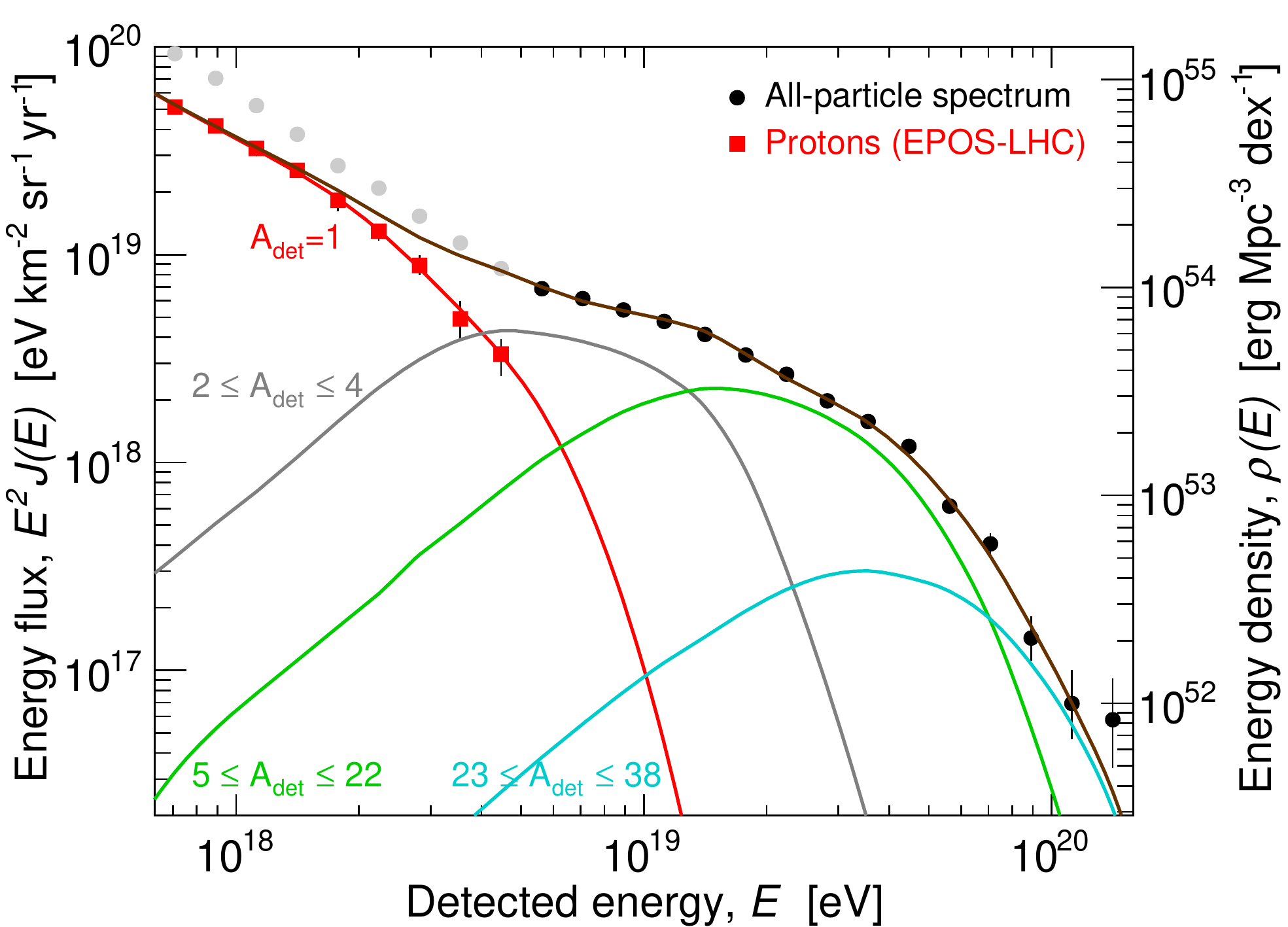}
\caption{Energy flux at Earth as a function of energy, as modeled by the best-fit parameters for the benchmark scenario. The all-particle spectrum and proton component are shown as black circles and red squares, respectively ~\citep{PierreAuger:2021hun}. Lighter points are not included in the fit. The best-fit components obtained for five detected mass groups are displayed with solid colored lines, as labeled in the Figure. The energy flux of the heaviest mass group, with detected mass number in 39--56, is below the range of interest.}
\label{fig:FitFlux}
\end{figure}

\begin{figure}[t]
\centering
\includegraphics[scale=.5]{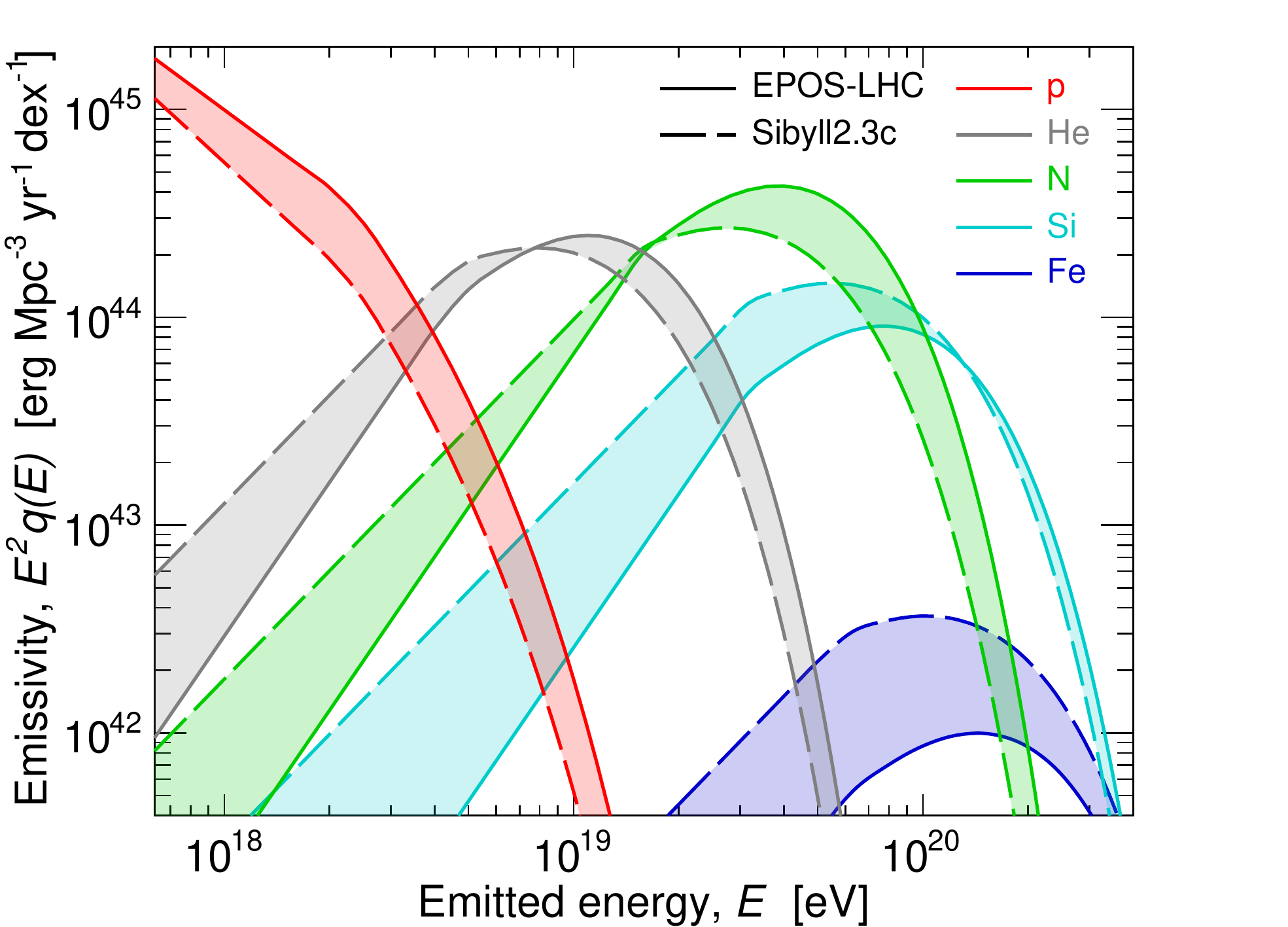}
\caption{Mass-dependent energy production rate at the sources as a function of energy, as constrained by the best-fit parameters for the benchmark scenario. The dashed lines illustrate a variation of the hadronic interaction model, with EPOS-LHC and Sibyll2.3c shown as solid and dashed lines, respectively.}
\label{fig:EProdRate}
\end{figure}

The balance determining the intensity of each component is reported in terms of energy production rates $\bar{\mathcal{L}}$ greater than 0.63\:EeV.
The result is displayed in Figure~\ref{fig:FitFlux}, which shows the contributions of each nuclear component to the observed energy flux and energy density. The energy production rates necessary at the sources to power the observed energy flux are shown as a function of energy in Figure~\ref{fig:EProdRate} for the various primary mass groups.
The solution is illustrated in Figure~\ref{fig:FitFlux},
While the contribution of nuclei is highest at ${\simeq}\: 5 \times 10^{44}\: {\rm erg}\: {\rm Mpc}^{-3}\: {\rm yr}^{-1}\: {\rm dex}^{-1}$, that of protons is increasing up to ${\simeq}\: 10^{45}\: {\rm erg}\: {\rm Mpc}^{-3}\: {\rm yr}^{-1}\: {\rm dex}^{-1}$ when going down in energy. Extrapolations of the results below 0.63\:EeV, on the other hand, are unsafe since the functional form employed in Equation~\eqref{eqn:qN}  may no longer hold depending on the features of the source environments that determine the nuclear cascade. The total energy production rate, integrating above 0.63\:EeV, is found to be $(10.8 \pm 0.4 )\times 10^{44}\: {\rm erg}\: {\rm Mpc}^{-3}\: {\rm yr}^{-1}$.

The reduced deviance is divided into three terms, according to Equation~\eqref{eqn:deviance}. The fit of the spectra leads to an acceptable fit ($D_J+D_{J_{\mathrm{p}}}=37.3$ for $N_J+N_{J_{\mathrm{p}}}=24$ points). As in~\cite{PierreAuger:2017pzq}, the $X_{\mathrm{max}}$ sector has a worse deviance (value $D_{X_{\mathrm{max}}}=199.5$ for $N_{X_{\mathrm{max}}}=109$ points). In general, the benchmark scenario has a fit quality  quite similar to those obtained by considering only data beyond 5\:EeV. Considering the variation of Equation~\eqref{eqn:qN} that consists in substituting $E_{\mathrm{max}}$ for $E_{\mathrm{max}}/2$, it was found that the main changes concern the spectral index of protons, which becomes $\gamma_{\mathrm{p}}\simeq 2.5$, while the deviance does not change between the two scenarios.

\section{Discussion} 
\label{sec:discussion}

The results help to explain the origin of protons below the ankle energy and the hard spectra of nuclei above this energy. The component of protons in this scenario is extragalactic in origin and exponentially suppressed above the ankle energy, while heavier nuclei slowly take over to the highest energies via a rigidity-dependent maximum-energy scenario. A softer spectral index for this population of light primaries reflects the fact that protons are not suppressed when they lose energy at the same rate as heavier nuclei. Remarkably, such behavior qualitatively corresponds to scenarios of in-source interactions in which abundant fluxes of neutrons originated by accelerated charged particles interacting with the bath of photons permeating the sources escape unimpeded from the electromagnetic fields.  Our conclusions provide a solid base for constraining the spectral indices $\gamma_{\rm p}$ and $\gamma_A$, as well as the energetics of the sources and the abundances of elements in their surroundings. There are two reasons for the constraints' robustness. First, by taking into account only the proton spectrum between 0.63~EeV and 5~EeV, we can rebuild the extragalactic component as directly and accurately as possible, with no "interference" from the expected  composition of the supposed end of the galactic component.
Second, by adding a local overdensity to the widely used uniform distribution of sources in a comoving volume, we can match the prediction with the data on a statistical basis characterized by $D/{\rm ndf}\simeq 237/125$ i.e. a reduced deviance on the order 
of that found in \cite{PierreAuger:2016use} for data strictly above the ankle. The study presented here significantly enhances the description of the data, particularly in the mass-composition
sector. Future observations will thus provide evidence to either confirm the role of in-source interactions in the interpretation of UHECR data, by detecting such a sub-dominant component below the ankle and its nature, either galactic or extra-galactic, with the upgraded instrumentation of the Pierre Auger Observatory in the next years~\citep{Castellina:2019irv}.

\end{document}